\documentclass[twocolumn,showpacs,preprintnumbers,amsmath,aps]{revtex4}
\usepackage{graphicx}
\usepackage{dcolumn}
\usepackage{bm}
\begin{document}
\newcommand{\kvec}{\mbox{{\scriptsize {\bf k}}}}
\def\eq#1{(\ref{#1})}
\def\fig#1{Fig.\hspace{1mm}\ref{#1}}
\def\tab#1{table\hspace{1mm}\ref{#1}}
\title{Superconductivity of calcium under the pressure at 120 GPa}
\author{R. Szcz{\c{e}}{\`s}niak, A.P. Durajski}
\affiliation{Institute of Physics, Cz{\c{e}}stochowa University of Technology, Al. Armii Krajowej 19, 42-200 Cz{\c{e}}stochowa, Poland}
\email{adurajski@wip.pcz.pl}
\date{\today} 
\begin{abstract}
The properties of the superconducting state in calcium under the pressure at $120$ GPa were analyzed. By using the imaginary axis Eliashberg equations it has been shown, that the Coulomb pseudopotential reaches the high value equal to $0.215$. In the considered case, the critical temperature is not properly described by the Allen-Dynes formula and it should be calculated with an use of the modified expression. In the paper the exact solutions of the Eliashberg equations on the real axis were also obtained. On this basis it was stated, that the effective potential of the electron-electron interaction is attractive for the frequencies lower or equal to the maximum phonon frequency. Then, the dimensionless parameter $2\Delta\left(0\right)/k_{B}T_{C}=4.10$ was calculated. In the last step it has been proven, that the ratio of the electron effective mass to the bare electron mass is high and reaches its maximum equal to $2.36$ for the critical temperature.
\end{abstract}
\maketitle
%
\section{Introduction}

\begin{figure}[t]%
\includegraphics*[scale=0.31]{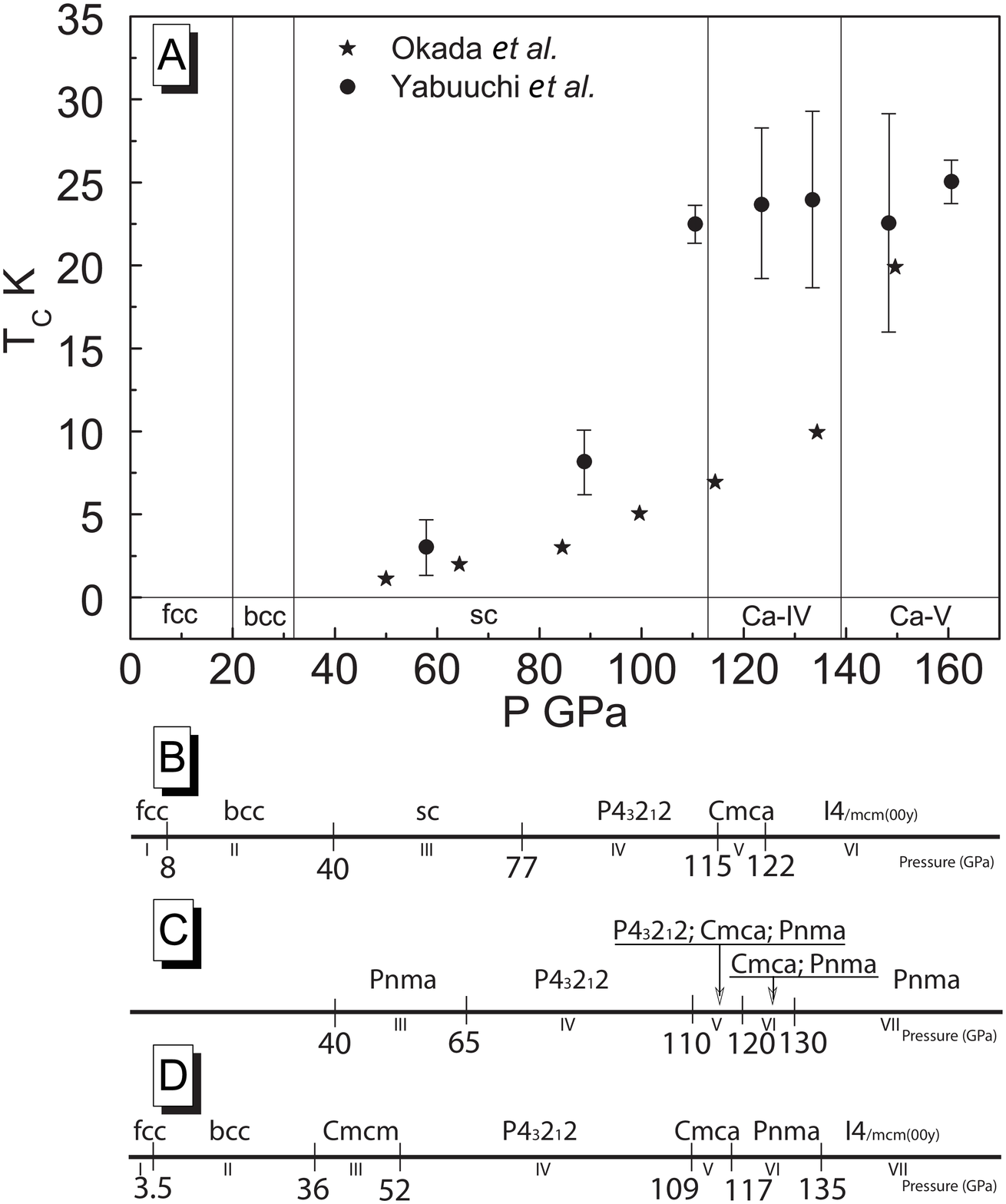}
\caption{
(A) The dependence of the critical temperature on the pressure for calcium. Experimental data:
 stars - S. Okada {\it et al.} \cite{Okada},
circles - T. Yabuuchi {\it et al.} \cite{Yabuuchi}. 
Additionally, we present a sequence of the successive structural transitions predicted by:
(A) - T. Yabuuchi {\it et al.} \cite{Yabuuchi} (see also the papers \cite{Olijnyk}-\cite{Nakamoto}),
(B) - S. Arapan {\it et al.} \cite{Arapan},
(C) - Z.P. Yin {\it et al.} \cite{Yin},
(D) - T. Ishikawa {\it et al.} \cite{Ishikawa}.}
\label{f1}
\end{figure}

The first mention of the pressure-induced superconducting state in calcium was given by Dunn and Bundy in 1981 \cite{Dunn}. In 1996 Okada {\it et al.} have obtained experimentally the dependence of the critical temperature ($T_{C}$) on the pressure ($p$) \cite{Okada}. The authors stated, that the critical temperature grows together with an increase of the pressure from the value of about $1$ K for $p = 50$ GPa to the value $\sim 20$ K for $150$ GPa. The results of Okada were verified in 2006 by Yabuuchi {\it et al.} \cite{Yabuuchi}. The significant discrepancy between new and old data was observed (see \fig{f1}). Most probably the difference between the experimental results comes from the use of the samples of distinct purity or from the different methods of the pressure measuring. The discussion of the considered problem the reader can find in the paper \cite{Yabuuchi}.

%
%

According to Yabuuchi {\it et al.} the superconducting state in calcium can be observed in the following crystal structures: {\it sc}, Ca-IV and Ca-V. In the structure {\it sc} the critical temperature quickly increases together with the growth of the pressure (from the value of about $3$ K for $p = 58$ GPa to the value $\sim 23$ K for the pressure equal to $113$ GPa). In the case of Ca-IV and Ca-V the increment of $T_{C}$ is much slower. Although, with the further compression to $161$ GPa the critical temperature increases to the value equal to $25$ K (the highest observed value of the critical temperature for the simple metals).

The issue of the $T_{C}$ value for the given pressure seems to be solved by Yabuuchi {\it et al.}. Unfortunately, the determination of the proper sequence of the structural transitions is still the tangled problem. On the one hand it is connected with the objective experimental difficulties, on the other hand the {\it ab initio} calculations are also burdened with the large uncertainty due to the small difference between the values of the enthalpy of the considered crystal structures \cite{Yin}.

Basing on the results presented in \cite{Yabuuchi}-\cite{Nakamoto} it can be stated, that in the normal conditions calcium crystallizes in {\it fcc} structure (\fig{f1} (A)). Under the influence of the pressure $20$ GPa the structural transition {\it fcc}-{\it bcc} was observed; above $32$ GPa calcium crystallizes in {\it sc} structure. The crystal structure {\it sc} is stable up to the pressure of about $113$ GPa, where another structural change (Ca-IV) was discovered. The last transition to Ca-V structure was observed for the pressure of $139$ GPa. It is worth mentioning, that the crystal structures of the draft symbols Ca-IV and Ca-V were experimentally classified by Fujihisa {\it et al.} in the paper \cite{Fujihisa}. In particular, the structure Ca-IV is identified with $P_{4_{1}2_{1}2}$ and Ca-V with {\it Cmca}.

Let us point the reader's attention toward the fact, that some recently published papers partially question above scheme of the structural transitions (see \fig{f1} (B)-(D)) \cite{Arapan}, \cite{Yin}, \cite{Ishikawa}. On the basis of the presented data it can be easily noticed, that the qualitative agreement between the obtained results exists only for the area of the low pressure's values (phases {\it fcc} and {\it bcc}). Above 
$\sim 36$ GPa the differences become significant. Thus at the present stage of the research, it is very difficult to specify which of the presented schemes is proper. 

From the physical point of view, the superconducting phase in calcium should have the most interesting properties at the area of high pressures, where
 $T_{C}>20$ K. For that reason, the thermodynamic parameters of superconducting state under the pressure at $120$ GPa were determined. In this case Yin {\it et al.} suggest, that structures: $P_{4_{3}2_{1}2}$, {\it Cmca} and {\it Pnma} are showing almost identical values of the enthalpy 
(\fig{f1} (C)) \cite{Yin}. The paper of Ishikawa {\it et al.} predicts the appearance of the crystal structure {\it Pnma} (\fig{f1} (D)) \cite{Ishikawa}.  

\section{THE ELIASHBERG FORMALISM}

In the paper, the thermodynamic parameters will be calculated in the strict way in the framework of the Eliashberg formalism \cite{Eliashberg}. In particular, the critical value of the Coulomb pseudopotential ($\mu^{*}_{C}$) and the parameters in the modified Allen-Dynes formula are going to be determined with an use of the Eliashberg equations defined on the imaginary axis \cite{Szczesniak1}-\cite{Szczesniak3}. The remain thermodynamic quantities (the order parameter and the electron effective mass) will be determined by using the Eliashberg equations in the mixed representation: the equations defined on both imaginary and real axis \cite{Marsiglio}.

\subsection{The Eliashberg equations on the imaginary axis}

The imaginary axis Eliashberg equations can be written in the following form \cite{Eliashberg}:
\begin{equation}
\label{r1}
\phi_{n}=\frac{\pi}{\beta}\sum_{m=-M}^{M}
\frac{\lambda\left(i\omega_{n}-i\omega_{m}\right)-\mu^{*}\theta\left(\omega_{c}-|\omega_{m}|\right)}
{\sqrt{\omega_m^2Z^{2}_{m}+\phi^{2}_{m}}}\phi_{m},
\end{equation}
\begin{equation}
\label{r2}
Z_{n}=1+\frac{1}{\omega_{n}}\frac{\pi}{\beta}\sum_{m=-M}^{M}
\frac{\lambda\left(i\omega_{n}-i\omega_{m}\right)}{\sqrt{\omega_m^2Z^{2}_{m}+\phi^{2}_{m}}}
\omega_{m}Z_{m},
\end{equation}
where the symbol $\phi_{n}\equiv\phi\left(i\omega_{n}\right)$ denotes the order parameter function and $Z_{n}\equiv Z\left(i\omega_{n}\right)$ is the wave function renormalization factor; $n$-th Matsubara frequency is given by the expression: 
$\omega_{n}\equiv \left(\pi / \beta\right)\left(2n-1\right)$, where $\beta\equiv\left(k_{B}T\right)^{-1}$ and $k_{B}$ is the Boltzmann constant. The order parameter is defined as the ratio: $\Delta_{n}\equiv \phi_{n}/Z_{n}$. The pairing kernel for the electron-phonon interaction is described with an use of the formula: 
\begin{equation}
\label{r3}
\lambda\left(z\right)\equiv 2\int_0^{\Omega_{\rm{max}}}d\Omega\frac{\Omega}{\Omega ^2-z^{2}}\alpha^{2}F\left(\Omega\right).
\end{equation}
The Eliashberg function for calcium ($\alpha^{2}F\left(\Omega\right)$) was determined in the paper \cite{Yin}; the maximum phonon frequency ($\Omega_{\rm{max}}$) is equal to $61.68$ meV.

In the framework of the Eliashberg formalism, the depairing interaction between the electrons is taken into consideration with an aid of the Coulomb pseudopotential $\mu^{*}$. The symbol $\theta$ denotes the Heaviside unit function and $\omega_{c}$ is the cut-off frequency; $\omega_{c}=3\Omega_{\rm{max}}$.

The Eliashberg equations were solved for $2200$ Matsubara frequencies ($M=1100$) with an use of the iteration method, described in the papers \cite{Szczesniak4} and \cite{Szczesniak5}. In the considered case the solutions of the Eliashberg equations are stable for $T\geq 5$ K.

\subsection{The critical value of the Coulomb pseudopotential}

\begin{figure}[t]%
\includegraphics*[scale=0.31]{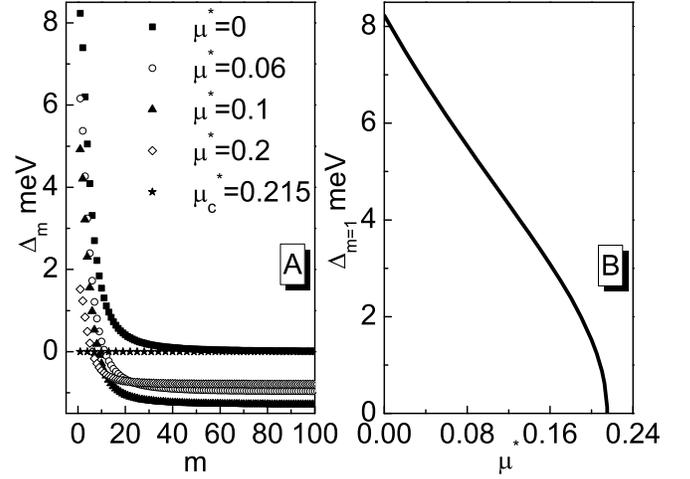}
\caption{
(A) The order parameter on the imaginary axis for the selected values of the Coulomb pseudopotential (the first $100$ values of $\Delta_{m}$ is presented). (B) The full dependence of the maximum value of the order parameter on the Coulomb pseudopotential.}
\label{f2} 
\end{figure}

The Coulomb pseudopotential is the second input parameter in the Eliashberg equations. In general, its critical value may be theoretically determined from the first principles, however such types of calculations are very complicated and obtained results are usually burdened with the large error. According to the above, the parameter $\mu_{C}^{*}$ will be selected in such way, that $T_{C}$ determined on the basis of the Eliashberg equations equals the experimental value of the critical temperature. In the paper $T_{C}=24$ K was assumed, in fair agreement with the results obtained by Yabuuchi (see \fig{f1}).

In order to succeed in above, the Eliashberg equations were solved in a strict way, for successive larger values of the Coulomb pseudopotential 
($T=T_{C}$). Achieved results are presented in \fig{f2} (A), where the dependence of the order parameter on the Matsubara frequencies is being plotted. It can be easily noticed, that together with the increase of $\mu^{*}$ the maximum value of the order parameter ($\Delta_{m=1}$) quickly decreases (\fig{f2} (B)). On the basis of the plotted results, it has been stated, that $\mu^{*}_{C}=0.215$.

Let us notice, that the critical value of the Coulomb pseudopotential determined with an use of the Eliashberg equations significantly differs from the value $\mu^{*}_{C}$ calculated with an aid of the classical formulas. In particular, the Allen-Dynes expression predicts: $\mu^{*}_{C}=0.168$ \cite{AllenDynes}; the estimations made on the basis of the McMillan formula are even more inaccurate ($\mu^{*}_{C}$=0.151) \cite{McMillan}. 

\subsection{The critical temperature}

%
\begin{figure}[t]%
\includegraphics*[scale=0.31]{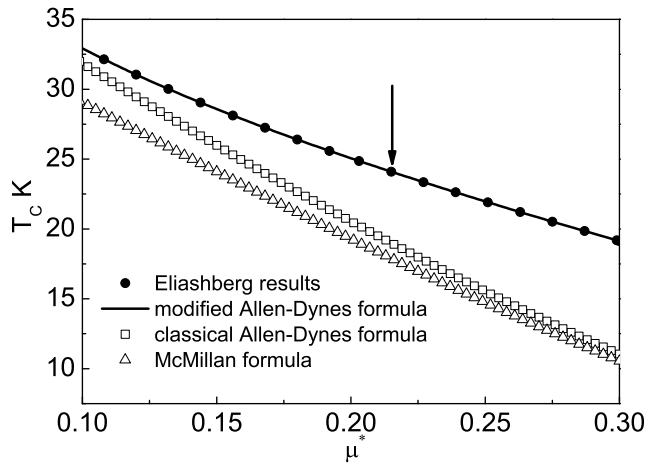}
\caption{
The value of the critical temperature as a function of the Coulomb pseudopotential. The filled circles represent the exact results obtained with an use of the Eliashberg equations. The solid line was achieved using the modified Allen-Dynes formula. The arrow shows the value $\mu_{C}^{*}=0.215$. The empty squares and triangles are related respectively to the classical Allen-Dynes expression and the McMillan formula.}
\label{f3} 
\end{figure}
%

In the previous paragraph we have shown that the Allen-Dynes or McMillan formula can not be used in the case of calcium. For that reason, the values of the parameters appearing in the classical Allen-Dynes formula were determined once again. In particular, we were basing on the $200$ exact values of $T_{C}\left(\mu^{*}\right)$ determined with an use of the Eliashberg equations. The following result was received: 
\begin{equation}
\label{r4}
k_{B}T_{C}=f_{1}f_{2}\frac{\omega_{\rm ln}}{1.45}\exp\left[\frac{-1.03\left(1+\lambda\right)}{\lambda-\mu^{*}\left(1+0.06\lambda\right)}\right],
\end{equation}
where the symbol $f_{1}$ ($f_{2}$) is the strong-coupling correction (shape correction) function:
\begin{equation}
\label{r5}
f_{1}\equiv\left[1+\left(\frac{\lambda}{\Lambda_{1}}\right)^{\frac{3}{2}}\right]^{\frac{1}{3}} \qquad {\rm and} \qquad
f_{2}\equiv 1+\frac{\left(\frac{\sqrt{\omega_{2}}}{\omega_{\rm{ln}}}-1\right)\lambda^{2}}{\lambda^{2}+\Lambda^{2}_{2}}.
\end{equation}
The parameter $\omega_{2}$ denotes the second moment of the normalized weight function: 
$\omega_{2}\equiv\frac{2}{\lambda}\int^{\Omega_{\rm{max}}}_{0}d\Omega\alpha^{2}F\left(\Omega\right)\Omega$.
The quantity $\omega_{{\rm ln}}$ is called the logarithmic phonon frequency and $\lambda$ is the electron-phonon coupling constant: 
$\omega_{{\rm ln}}\equiv \exp\left[\frac{2}{\lambda}
\int^{\Omega_{\rm{max}}}_{0}d\Omega\frac{\alpha^{2}F\left(\Omega\right)}
{\Omega}\ln\left(\Omega\right)\right]$ and $\lambda\equiv 2\int^{\Omega_{\rm{max}}}_{0}d\Omega\frac{\alpha^{2}F\left(\Omega\right)}{\Omega}$.
For calcium it was achieved: $\sqrt{\omega_{2}}=29.81$ $\rm{meV}$, $\omega_{{\rm ln}}=25.34$ meV and $\lambda=1.3$. In the paper we have changed also the parametrization of the functions $\Lambda_{1}$ and $\Lambda_{2}$. We have obtained:
\begin{equation}
\label{r6}
\Lambda_{1}\equiv 2\left(1+4.7\mu^{*}\right)
\end{equation}
and 
\begin{equation}
\label{r7}
\Lambda_{2}\equiv -0.085\left(1-150\mu^{*}\right)\left(\frac{\sqrt{\omega_{2}}}{\omega_{\rm{ln}}}\right).
\end{equation}

In \fig{f3} we present a dependence of the critical temperature on the Coulomb pseudopotential. The results were obtained with an use of the Eliashberg equations, the modified and classical Allen-Dynes formula. Additionally, the values of $T_{C}$ calculated on the basis of the McMillan formula ($f_{1}=f_{2}=1$) are also plotted. It can be easily noticed, that the modified Allen-Dynes formula allows to very precisely reproduce the results of the Eliashberg theory in the whole considered range of $\mu^{*}$.

\subsection{The dependence of $\Delta_{m}$ and $Z_{m}$ on the temperature}

%
\begin{figure}[t]%
\includegraphics*[scale=0.31]{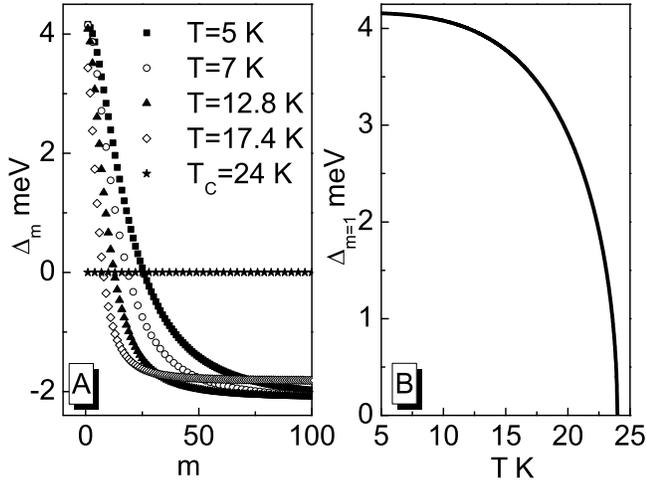}
\caption{(A) The order parameter on the imaginary axis for the selected temperatures; first $100$ values of $\Delta_{m}$ is presented. (B) The dependence of the maximum value of the order parameter on the temperature. The function can be parameterized with an use of the expression:  $\Delta_{m=1}\left(T\right)=\Delta_{m=1}\left(0\right)\sqrt{1-\left(\frac{T}{T_{C}}\right)^{\beta}}$, where $\Delta_{m=1}\left(0\right)=4.16$ meV and $\beta=3.7$.}
\label{f4}
\end{figure}
%
\begin{figure}[t]%
\includegraphics*[scale=0.31]{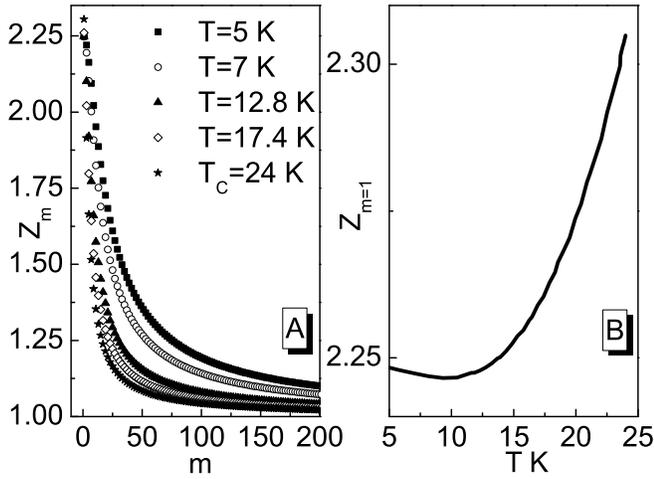}
\caption{(A) The wave function renormalization factor on the imaginary axis for the selected temperatures; first $200$ values of $Z_{m}$ is presented. (B) The dependence of the maximum value of the wave function renormalization factor on the temperature.}
\label{f5}
\end{figure}
%

In the paragraph there are presented solutions of the Eliashberg equations on the imaginary axis for the temperature range from $5$ K to $T_{C}$. In particular, in \fig{f4} (A) we have plotted the dependence of the order parameter's values on the successive Matsubara frequencies. It can be easily noticed, that with the growth of the temperature the maximum of the order parameter's function ($\Delta_{m=1}$) is decreasing. Additionally, the 
half-width of the function becomes successively smaller. The last property means, that together with the temperature's growth less successive Matsubara frequencies commit a relevant contribution to the Eliashberg equations. Let us notice, that the dependence of the order parameter on temperature most conveniently can be traced after plotting function $\Delta_{m=1}\left(T\right)$ (see \fig{f4} (B)). The obtained result proves, that the maximum value of the order parameter becomes saturated for the temperature equal to $5$ K. For that reason the order parameter for $T=5$ K can be used in the calculations of the energy gap at the temperature of zero Kelvin.

The second solution of the Eliashberg equations has been shown in \fig{f5} (A). Also in this case, for the increasing values of the Matsubara frequencies the values of $Z_{m}$ are decreasing. However, in comparison with the order parameter, the decrease is much slower. The dependence of the maximum value of the wave function renormalization factor on the temperature looks completely different than for $\Delta_{m=1}$ (\fig{f5} (B)). It can be seen, that together with the growth of temperature $Z_{m=1}$ has the shallow minimum and then slightly increases. 

\subsection{The Eliashberg equations in the mixed representation}

In order to exactly calculate the value of the energy gap at the temperature of zero Kelvin and electron effective mass
one should have the solutions of Eliashberg equations on the real axis. From the mathematical point of view the determination of such type function is a very complicated issue. Usually to do that, one uses the solutions on the imaginary axis and then analytically continues them on the real axis \cite{AnalKon}. However, the described procedure has a serious drawback e.g. the obtained functions are stable only for the low values of the frequencies (much lower than the maximum phonon frequency). For that reason, we have used a method of calculation which is deprived of the weak point mentioned above \cite{Marsiglio}. In particular, the approach is based on the transformation of the Eliashberg equations to the mixed representation, in which the order parameter function and the wave function renormalization factor are defined on both real and imaginary axis. In the considered case the Eliashberg set takes the following form:
%
\begin{widetext}
\begin{eqnarray}
\label{r8}
\phi\left(\omega+i\delta\right)&=&
                                  \frac{\pi}{\beta}\sum_{m=-M}^{M}
                                  \left[\lambda\left(\omega-i\omega_{m}\right)-\mu^{*}\theta\left(\omega_{c}-|\omega_{m}|\right)\right]
                                  \frac{\phi_{m}}
                                  {\sqrt{\omega_m^2Z^{2}_{m}+\phi^{2}_{m}}}\\ \nonumber
                              &+& i\pi\int_{0}^{+\infty}d\omega^{'}\alpha^{2}F\left(\omega^{'}\right)
                                  \left[\left[N\left(\omega^{'}\right)+f\left(\omega^{'}-\omega\right)\right]
                                  \frac{\phi\left(\omega-\omega^{'}+i\delta\right)}
                                  {\sqrt{\left(\omega-\omega^{'}\right)^{2}Z^{2}\left(\omega-\omega^{'}+i\delta\right)
                                  -\phi^{2}\left(\omega-\omega^{'}+i\delta\right)}}\right]\\ \nonumber
                              &+& i\pi\int_{0}^{+\infty}d\omega^{'}\alpha^{2}F\left(\omega^{'}\right)
                                  \left[\left[N\left(\omega^{'}\right)+f\left(\omega^{'}+\omega\right)\right]
                                  \frac{\phi\left(\omega+\omega^{'}+i\delta\right)}
                                  {\sqrt{\left(\omega+\omega^{'}\right)^{2}Z^{2}\left(\omega+\omega^{'}+i\delta\right)
                                  -\phi^{2}\left(\omega+\omega^{'}+i\delta\right)}}\right]
\end{eqnarray}
and
\begin{eqnarray}
\label{r9}
Z\left(\omega+i\delta\right)&=&
                                  1+\frac{i}{\omega}\frac{\pi}{\beta}\sum_{m=-M}^{M}
                                  \lambda\left(\omega-i\omega_{m}\right)
                                  \frac{\omega_{m}Z_{m}}
                                  {\sqrt{\omega_m^2Z^{2}_{m}+\phi^{2}_{m}}}\\ \nonumber
                              &+&\frac{i\pi}{\omega}\int_{0}^{+\infty}d\omega^{'}\alpha^{2}F\left(\omega^{'}\right)
                                  \left[\left[N\left(\omega^{'}\right)+f\left(\omega^{'}-\omega\right)\right]
                                  \frac{\left(\omega-\omega^{'}\right)Z\left(\omega-\omega^{'}+i\delta\right)}
                                  {\sqrt{\left(\omega-\omega^{'}\right)^{2}Z^{2}\left(\omega-\omega^{'}+i\delta\right)
                                  -\phi^{2}\left(\omega-\omega^{'}+i\delta\right)}}\right]\\ \nonumber
                              &+&\frac{i\pi}{\omega}\int_{0}^{+\infty}d\omega^{'}\alpha^{2}F\left(\omega^{'}\right)
                                  \left[\left[N\left(\omega^{'}\right)+f\left(\omega^{'}+\omega\right)\right]
                                  \frac{\left(\omega+\omega^{'}\right)Z\left(\omega+\omega^{'}+i\delta\right)}
                                  {\sqrt{\left(\omega+\omega^{'}\right)^{2}Z^{2}\left(\omega+\omega^{'}+i\delta\right)
                                  -\phi^{2}\left(\omega+\omega^{'}+i\delta\right)}}\right], 
\end{eqnarray}
\end{widetext}
%
where symbols $N\left(\omega\right)$ and $f\left(\omega\right)$ denote the statistical functions of bosons and fermions respectively.

\subsection{The order parameter on the real axis}

%
\begin{figure}[t]%
\includegraphics*[scale=0.31]{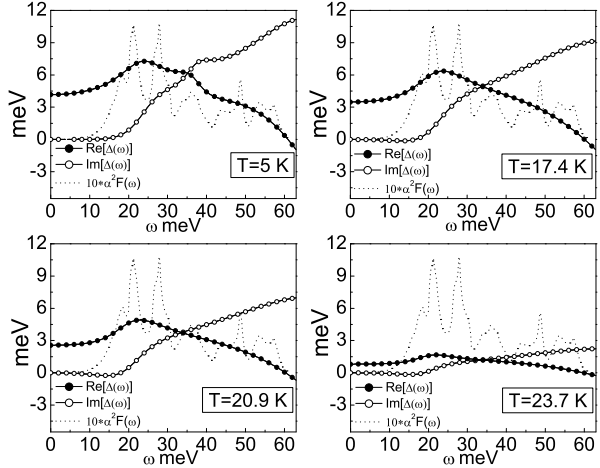}
\caption{
The dependence of the real and imaginary part of the order parameter on the frequency for the selected temperatures. The rescaled Eliashberg function is also plotted.} 
\label{f6}
\end{figure}
%
\begin{figure}[t]%
\includegraphics*[scale=0.31]{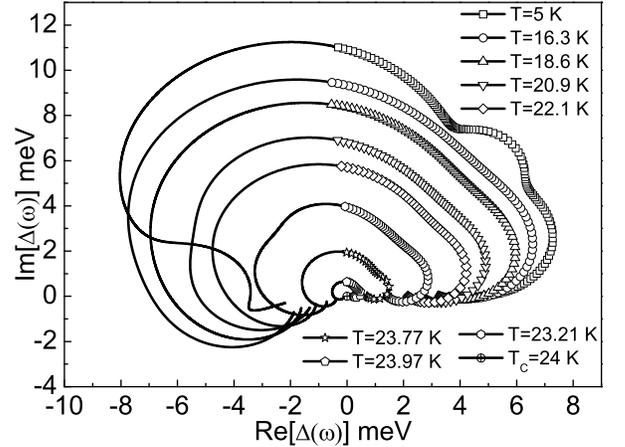}
\caption{
The values of $\Delta\left(\omega\right)$ on the complex plane for the selected temperatures.}
\label{f7}
\end{figure}
%

The Eliashberg equations in the mixed representation were solved for identical temperatures like the equations on the imaginary axis. In \fig{f6} we have presented Re$\left[\Delta\left(\omega\right)\right]$ and Im$\left[\Delta\left(\omega\right)\right]$ for the range of the frequencies from $0$ to $\Omega_{\rm max}$ and selected temperatures. Additionally, the form of the Eliashberg function is also plotted. It can be easily noticed, that for the low frequencies the non-zero is only the real part of the order parameter. From the physical point of view it means the lack of the damping effects, which are described by the imaginary part of the order parameter \cite{Varelogiannis}. For higher frequencies the functions Re$\left[\Delta\left(\omega\right)\right]$ and Im$\left[\Delta\left(\omega\right)\right]$ are characterized by the complicated course which is very plainly correlated with the shape of the Eliashberg function. 

The full dependence of the order parameter on the temperature for the range of the frequencies from $0$ to $\omega_{c}$ can be traced in the most convenient way on the complex plane (see \fig{f7}). On the basis of the presented results it can be seen, that the values of the order parameter are forming the characteristic spirals with radius that decreases together with the temperature growth. In \fig{f7} the lines with open symbols represent the solutions for $\omega\in\left<0,\Omega_{\rm max}\right>$; whereas solid lines correspond to the solutions for $\omega\in\left(\Omega_{\rm max}, \omega_{c}\right>$. In the first case the condition Re$\left[\Delta\left(\omega\right)\right]>0$ is fulfilled with a good approximation and in the second case Re$\left[\Delta\left(\omega\right)\right]<0$ occurs. Let us notice, that the real part of the order parameter is connected with the effective pairing potential for the electron-electron interaction \cite{Varelogiannis}. Thus, the obtained result means, that only in the range of the frequencies for which the Eliashberg function is defined, the effective potential for the electron-electron interaction is attractive.

Below on the basis of the solution for $T=5$ K we have calculated the value of the order parameter at the temperature of zero Kelvin ($\Delta\left(0\right)$). In particular to achieve that, the following equation was used:   
\begin{equation}
\label{r10}
\Delta\left(T\right)={\rm Re}\left[\Delta\left(\omega=\Delta\left(T\right)\right)\right].
\end{equation}
As a result it was obtained: $\Delta(0)=4.24$ meV. The familiarity with the energy gap ($2\Delta\left(0\right)$) allows to determine the dimensionless ratio $R_{1}\equiv\frac{2\Delta\left(0\right)}{k_{B}T_{C}}$, which for the calcium is equal to $4.10$. Let us notice, that in the framework of the BCS theory the parameter $R_{1}$ takes considerably smaller value that is equal to $3.53$ \cite{BCS}.

\subsection{The electron effective mass}

%
\begin{figure}[t]%
\includegraphics*[scale=0.31]{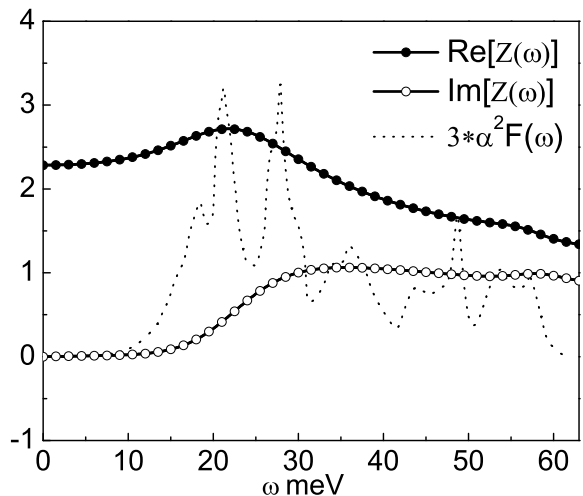}
\caption{
The dependence of the real and imaginary part of the wave function renormalization factor on the frequency for $T = 17.4$ K. The rescaled Eliashberg function is also plotted.}
\label{f8}
\end{figure}
%
\begin{figure}[t]%
\includegraphics*[scale=0.31]{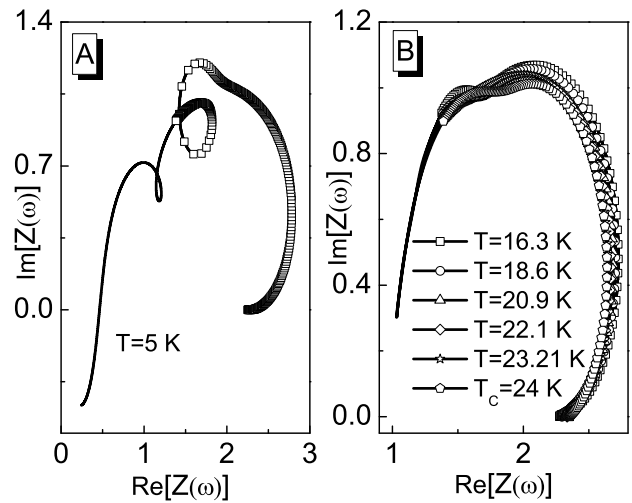}
\caption{
The values of $Z\left(\omega\right)$ on the complex plane for the selected temperatures. The lines with open symbols are for $\omega\in\left<0,\Omega_{\rm max}\right>$; the solid lines correspond to $\omega\in\left(\Omega_{\rm max},\omega_{c}\right>$. (A) Graph for $T=5$ K. (B) The courses in the area of the "high" temperatures.}
\label{f9}
\end{figure}
%

In \fig{f8} for selected value of the temperature the functions Re$\left[Z\left(\omega\right)\right]$ and Im$\left[Z\left(\omega\right)\right]$ are presented ($\omega\in\left<0,\Omega_{\rm max}\right>$). On the basis of the plotted courses it can be easily noticed, that for the low values of $\omega$ non-zero is only the real part of the wave function renormalization factor; this result is directly correlated with the order parameter's behavior. For higher frequencies the shape of the real and imaginary part of $Z\left(\omega\right)$ is also plainly dependent on the Eliashberg function's form.

The values of the wave function renormalization factor on the complex plane for the selected values of the temperature is presented in \fig{f9}. In particular, in \fig{f9} (A) we have shown an exemplary low-temperature solution which is characterized by a very complicated course; whereas in \fig{f9} (B) we have placed a few selected "high"-temperature solutions, which are presenting quite similar curves.    

In the framework of the Eliashberg formalism the real part of the wave function renormalization factor enables the determination of the electron effective mass ($m^{*}_{e}$). In particular, the ratio of $m^{*}_{e}$ to the bare electron mass ($m_{e}$) is given by: 
$m^{*}_{e}/m_{e}={\rm Re}\left[Z\left(0\right)\right]$. Let us notice that Re$\left[Z\left(0\right)\right]$ takes the highest value for $T=T_{C}$ (similarly as $Z_{m=1}$). Thus, $\left[m^{*}_{e}\right]_{{\rm max}}$ is equal to $2.36m_{e}$.

\section{Summary}

The basic thermodynamic quantities for the superconducting phase, that induces in calcium under the influence of the pressure $120$ GPa, were determined. All calculations were conducted in the framework of the Eliashberg formalism. In the first step it was shown, that the Coulomb pseudopotential takes very high value equal to $0.215$. Next it was stated, that the classical Allen-Dynes formula significantly underestimates the critical temperature. According to the above, the parameters in the Allen-Dynes formula were determined once again adapting the formula to the strict results predicted by the Eliashberg equations. The dimensionless parameter $R_{1}=4.10$ was also determined. The obtained result means, that the value of the energy gap at the temperature of zero Kelvin considerably exceeds the value predicted by the BCS theory. In the last step, it has been stated that the electron effective mass takes the highest value equal to $2.36m_{e}$ for $T=T_{C}$. 

The method of analysis based on the Eliashberg equations in the mixed representation allows to determine the dependence of the order parameter on the frequency for the range from $0$ to $\omega_{c}$. On the basis of the conducted calculations it was stated, that the values of $\Delta\left(\omega\right)$ on the complex plane lie on the characteristic spirals with the radius decreasing together with the temperature growth. The important notification is related to the fact, that for $\omega\in\left<0,\Omega_{\rm max}\right>$ the effective potential of the 
electron-electron interaction is attractive, while for $\omega\in\left(\Omega_{\rm max},\omega_{c}\right>$ is repulsive. 

\begin{acknowledgments}
The authors wish to thank Prof. K. Dzili{\'n}ski for providing excellent working conditions and the financial support. We also thank Mr M.W. Jarosik and Mr D. Szcz{\c{e}}{\'s}niak for the productive scientific discussion, that improved the quality of the presented paper. All numerical calculations were based on the Eliashberg function sent to us by: Prof. W.E. Pickett and Dr Z.P. Yin for whom we are very thankful.
\end{acknowledgments}
%

%
\end{document}